\documentclass[aps,pre,twocolumn,groupedaddress,showpacs]{revtex4}
\usepackage{latexsym}
\usepackage{amssymb}
\usepackage{graphicx}
\newcommand{\W}{8cm}
\begin{document}


\title{Single fiber transport  by a fluid flow in a fracture with rough walls: influence of the fluid rheology}



\author{Maria Veronica D'Angelo$^{1}$}\email{vdangelo@fi.uba.ar}
\author{Harold Auradou$^1$}\email{auradou@fast.u-psud.fr}
\author{Guillemette Picard$^2$}\email{gpicard@slb.com}
\author{Martin E. Poitzsch$^2$}\email{poitzsch1@slb.com}
\author{Jean-Pierre Hulin$^1$}\email{hulin@fast.u-psud.fr}

\affiliation{$^1$Univ Pierre et Marie Curie-Paris6, Univ Paris-Sud, CNRS, F-91405.
  Lab FAST, Bat 502, Campus Univ, Orsay, F-91405, France.\\$^2$Schlumberger-Doll Research, $1$ Hampshire Street, Cambridge, MA $02139$, U.S.A.}



\date{\today}

\begin{abstract}
The possible transport of fibers by fluid flow in fractures is investigated experimentally in  transparent models
using flexible polyester thread  (mean diameter $280 \mu\mathrm{m}$) and Newtonian and shear thinning fluids.
In the case of smooth parallel walls, fibers of finite length $\ell = 20-150\, \mathrm{mm}$ move at a constant
velocity of the order of the maximum fluid velocity in the aperture.
In contrast, for fibers lying initially at the inlet side of the model and dragged by the flow inside it,
the velocity increases  with the
depth of penetration (this results from the lower velocity - and drag - in the inlet part).
In both cases, the  friction of the fiber with the smooth walls is weak.
For  rough self-affine walls and a continuous gradient of the local mean aperture  transverse to the flow,
transport of the fibers by a water flow is only possible in the region of larger  aperture
 ($\bar{a} \gtrsim 1.1 \mathrm{mm}$) and is of ``stop and go'' type at low velocities.  Time dependent distorsions of the fiber are also often observed.
When water is replaced by a shear thinning polymer solution,  the fibers
move faster and  continuously in high aperture regions and their friction with the walls is reduced.
Fiber transport becomes also possible in narrower regions where irreversible pinning occurred for water.
In a third rough model with no global aperture  gradient but with rough walls and a  channelization parallel
 to the mean flow, fiber transport was only possible in shear-thinning flows and pinning and entanglement
 effects were studied.
\end{abstract}


\maketitle

\section{Introduction}
\label{intro}
Fiber transport by flowing fluids is of interest in many areas of
physics, biology and engineering: examples
include the transport of paper pulp~\cite{Stockie1998},
the manufacturing of fiber-reinforced
composites~\cite{Yasuda2002}, the rheology of
biological polymers~\cite{Lagomarsino2005} and the motility
of micro-organisms~\cite{Lowe2003,Purcell1977}. Recently, also, using long optical
fibers has been suggested as a method to realize distributed in-situ measurements (temperature for
instance) on  natural water flows~\cite{Selker2006}:  this raises the problem of the
possible transport of the fiber by the moving fluid which often
occurs in geometries confined by solid walls.
More precisely,  little work has been
devoted to flow channels with rough walls  such as fractures of natural
rocks and of materials of industrial interest in civil, environmental and petroleum
engineering. In that case,  the interactions
of the fibers with the walls are particularly important and may lead to blockage of the motion
of the fibers and/or to clogging of the channels.
In addition to their practical applications, these processes raise fundamental questions regarding
the motion of flexible solid bodies in complex flow fields.

The objective of the present experimental study is to understand the transport of fibers by the
flow and to investigate the role of the flow geometry and fluid rheology.
Here, single  fractures are modeled by the space (saturated by a flowing fluid) between
either two parallel plane walls or two  complementary rough self-affine walls with a relative
shear displacement from their contact position. This latter configuration allows one to
reproduce preferential flow channels~\cite{NAS,Adler1999} which are a widespread
feature of natural  fractures and influence strongly their transport properties.
The walls of the fracture are transparent to allow for optical observations of the motion of the fibers.

In addition to hydrodynamic forces on the fibers due to the relative velocity with the fluid,
their motion and deformation is influenced by different effects.
A first one is the interaction forces with the walls: they are
particularly important when the diameter of the fibers
is comparable to the local channel aperture or  when they are close to one of the
walls~\cite{Sugihara-Seki1993,Petrich1998}.
Second, tension forces reflecting the mechanical cohesion of the fiber are present all along
its length so that the motion of each region influences the other ones. Tension and
hydrodynamic forces add-up and their  spatial variations in shear flows or flows
with curved streamlines  deform the fibers (although their length
remains constant): this may  finally lead to entanglement and blockage.
Next, these deformations are opposed by elastic forces reflecting the
non-zero stiffness of the fibers:  their relative magnitude with respect to the
hydrodynamic forces  is a particularly important parameter of the
problem~\cite{Forgac1959}. On the one hand, very flexible fibers would seem to
be able to follow the streamlines but, on the other hand, loops may appear easily and lead to trapping
in narrow zones~\cite{Forgac1959b}.
Another key element is  the rheology of the fluids  which  influences strongly
the hydrodynamic forces on slender bodies, even in simple shear flows~\cite{Leal1975}.
Finally, inertia (particularly in regions of large spatial flow velocity variations)
and gravity  may also be of importance by inducing motions transverse to the
flow and towards the walls.

In the following, the feasibility of fiber transport and its dependence on the mechanisms discussed above
are studied in three different model fractures.
A first one has smooth walls and is used as a reference case and the two others have rough walls with self
affine geometries. For  one of the rough fractures ($F3$), the mean planes of the walls
are parallel. For the other one ($F2$), they have a small angle resulting in a non-zero
transverse gradient of the mean aperture: this wedge shape mimicks the edge of many
natural fractures.
In this work, the influence of the geometry of the fibers on their transport is analyzed by comparing
the  motion of  finite length segments (still with an aspect ratio greater than $200$)
and of continuous threads injected at the inlet of the fracture.
The influence of deformations  due to flow velocity gradients could be investigated
by using flexible thin fibers made of polyester thread.
Special attention has been brought to the mechanisms of
pinning during the motion of the fibers and of possible depinning after enough elastic
energy has  been
accumulated (see sections \ref{sec:F2W} and \ref{sec:F2NN}). Finally, the
influence of the fluid rheology has been studied by comparing fiber transport
by Newtonian and shear thinning fluids.
\section{Experimental set-up and procedure} \label{sec:exp}
\subsection{Fiber characteristics}
\label{fiber}
The experimental fibers  are prepared from commercial polyester thread used
for needlework and made of
two strands twisted together. The section is not circular so that the diameter varies
between $220$ and  $340\ \mu m$.
The specific mass of the
fiber is $\rho=1.8 \pm 0.1 \times 10^{-3} \, kg/m^3$. Its bending stiffness
$EI$ (ratio of the applied bending momentum by the curvature)
 is of the order of $10^{-9}\, kg.m^3/s^2$ (a value similar to that reported
 in ref.~\cite{habibi07} for a comparable material). This value
 has been estimated by measuring the deflection under its own weight
of a horizontal fiber segment attached at one end~\cite{Landau}.

The choice of these multifilament fibers reflects a trade-off between several
requirements. The fiber diameter is large enough to be clearly visible and its flexibility is both
high enough to allow for deformations by the fluid velocity gradients and low enough so that
coils do not build up too easily (see introduction). We verified indeed that monofilament wires
of similar (and even smaller) diameters were too rigid and retain often, in addition, a
permanent curvature.Ó

Both segments of fiber and continuous fibers were used in the present work.
The segments are cut out of polyester thread with a length
$20\, \mathrm{mm} \le \ell \le 150\, \mathrm{mm}$.
The continuous fibers are also cut out of a spooled  thread but with
 a length larger than that of the fracture: they are progressively injected from above into
 the fracture under zero applied tension conditions.
 A specific procedure is used to avoid blunting the ends of the fibers and
 the fibers are carefully saturated with liquid prior to the experiments.
\subsection{Model fractures}
\label{model}
\begin{figure}[h!]
\noindent\includegraphics[width=\W]{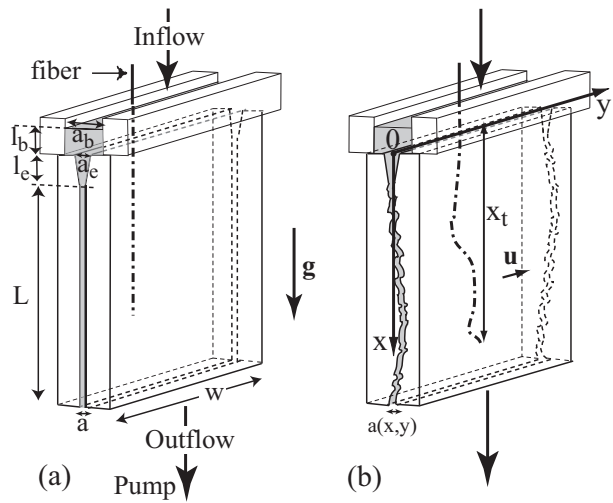}
\caption{Schematic view of the experimental models.  (a) fracture
with flat parallel walls - (b) fracture with complementary
self-affine walls with a relative displacement $\vec{u}$. For all cells : $w=90\ \mathrm{mm}$, $a_b =20\ \mathrm{mm}$,
$a_e = 5\ \mathrm{mm}$, $l_e=52\ \mathrm{mm}$, $l_b \sim 20\ \mathrm{mm}$
and $L=288\ \mathrm{mm}$. The mean flow is vertical and parallel to $x$} \label{fig:setup}
\end{figure}
The fracture models are manufactured with the same technique as in Ref.~\cite{Boschan2006}
by carving two parallelepipedic plexiglas blocks using a computer controlled milling
machine. The two blocks are then clamped together in a preset position determined by
the geometry of the sides of the
block: these act as spacers leaving a controlled interval  between the surfaces for the
 fluid flow (Fig.~\ref{fig:setup}).
This procedure allows for the realization of model fractures  with rough (or smooth)
walls of arbitrary geometries and   three of them  have been used.

The first sample (refered to as $F1$) has smooth parallel plane walls separated by a
fixed distance
 $a(x,y)=0.65\ \mathrm{mm}$ (Fig.~\ref{fig:setup}a).
The two other models have complementary rough walls with a self affine geometry
of characteristic Hurst exponent  $H = 0.8$: this value is similar to that measured on
granite fractures~\cite{Boffa1998} and in various other materials~\cite{Bouchaud2003}.
The amplitude of the roughness, represented by the difference between  the heights
at  the highest and lowest points of the surface is equal to $26\ \mathrm{mm}$: this
corresponds
also to values measured on  natural rock fractures~\cite{Poon1992}. For these
two models, the wall surfaces
are perfectly complementary: in order to generate aperture fluctuations, they
are positioned with a small relative shear displacement $\vec{u} = 0.65\ \mathrm{mm}$
in the direction  $y$ normal to the flow (Fig.~\ref{fig:setup}b).

The resulting aperture maps are displayed for the two models
in Figures~\ref{fig:fiberwedge} and \ref{fig:fiberparallel} using a color code. The bottom curves represents the
transverse profile in the $y$ direction of the average along $x$ of the aperture.
In model $F2$,
 there is a small angle between the mean plane of the surfaces mimicking the
 edge of natural  fractures.    As a result, the average $\overline{a}$ of the aperture along the direction $x$
 of the flow
 decreases with the coordinate $y$ from $\overline{a(y=0)}=1.1\ \mathrm{mm}$ to
 $\overline{a(y=w)}=0.8\ \mathrm{mm}$  with a global mean value $\bar{a} = 0.91\ \mathrm{mm}$
 (Fig.~\ref{fig:fiberwedge}c); $w$ is the width of the cell and $y = 0$ and $y = w$
correspond to the sides of  the experimental region. Note that the wedge shape of
the model results in a  global color gradient
in the direction $y$ in  the maps of Fig.~\ref{fig:fiberwedge}a-b.
The root mean square amplitude $\sigma_{a}(y)=(<(\overline{a(y)}-a(x,y))^2>_x)^{1/2}$
of the fluctuations of the aperture in the direction $x$ with respect to the mean value
$\overline{a(y)}=<a(x,y)>_x$ is equal to  $\sigma_a(y) =0.122\ \mathrm{mm}$ and independent
of $y$. Therefore, the fluctuations remain of constant absolute amplitude while the mean aperture varies
with $y$.

 In model $F3$, the mean planes of the fracture walls are
 parallel with a distance $a = 0.65\ \mathrm{mm}$ comparable to the value for model
 $F1$. The local fluctuations of the aperture are of the same order of magnitude as
 the deviations from the linear trend  in model $F2$  (Fig.~\ref{fig:fiberparallel}c).
 No global trend in the color shade at the scale of the global size of the fracture is visible in the map of  Fig.~\ref{fig:fiberparallel}a-b.
 Comparing experiments realized with the two systems  allows one to
 discriminate between the influence of the roughness and that of the global transverse aperture gradient.
\subsection{Experimental set-up and procedure}
\label{setup}
The model fractures are held  vertically; liquid is sucked uniformly at the bottom
side at a constant flow rate and reinjected at the top into an open bath of area
 $5 \times 100\ \mathrm{mm}$ and depth $l_b \sim 10 \, \mathrm{mm}$ covering the inlet of the model. This design allows one to
 introduce  the fibers in the bath through its open surface and then into the fracture.
 In order to make the injection of the fibers easier, the upper section has a funnel-like ``Y'' shape, {\it i.e.}
  the mean aperture of the fracture increases with height
 in the top  $52 \, \mathrm{mm}$ of the model (distance $l_e$ in Figs.~\ref{fig:setup}a-b).

The model is illuminated from behind by a light panel and a digital
camera provides images with $1024 \times 768$ pixels at a rate
of $30$ frames per second and with an exposure time of $1/300\ \mathrm{s}$.
The length of the field of view is $150\ \mathrm{mm}$ in the vertical direction parallel
 to the flow and the top of the aperture maps corresponds to a distance of $110\ \mathrm{mm}$
 from the inlet of the fracture.

On each picture, the location and geometry of the fiber is determined by a binary
thresholding technique. For most experiments, except in some cases in which the
fiber gets pinned, its distortions are moderate  and it remains overall aligned with the
direction of the flow. Therefore, at each distance $x$ from the inlet, there is only one point along
the fiber centerline, and, therefore, only one value of $y$.
>From these data,
spatiotemporal-diagrams like those displayed in Figs.~\ref{fig:spatio}a-b are obtained.
 The vertical scale corresponds to the vertical distance $x$, the horizontal
one to time and the color code corresponds to the distance $y$ at the corresponding value of $x$.
Qualitatively, these diagrams provide information on the global motion of the fibers, on their
rotation and on their deformation. For instance, diagrams with uniform colors on horizontal lines
 correspond to a pure translation of the fiber in the direction parallel to the mean flow;
in contrast, color variations of these lines imply lateral motions.
Quantitatively, the velocity of segments of fibers with a finite length
is take equal the variation between two successive images of the vertical  distance $x_m$
of their center of mass  to the inlet side of the model.
For continuous fibers, one uses instead the variation of the distance $x_t$ of the tip of the fiber to the inlet.  In both cases,
the vertical velocity component  of the center of mass (respectively tip) of the fiber will be referred to as $V_f$.
Continuous fibers are more easily retrieved after the experiments than fiber segments when trapping occurs: for that reason, most
experiments in rough fractures have been realized with continuous fibers  in order to obtain more data.
\subsection{Characteristics of the fluids}
\label{fluids}
The Newtonian fluid used in the experiments is high
purity water (Millipore - Milli-Q grade) with a specific mass
$\rho = 10^3\ \mathrm{kg/m^3}$ and a dynamic viscosity  $\mu_\infty \simeq 1 \
\mathrm{mPa.s}$. The shear thinning fluid is a solution
of high molecular weight Scleroglucan (Sanofi Bioindustries) in water at
a concentration  $C_p = 1000\ \mathrm{ppm}$.
The variation of the effective viscosity $\mu$ with the shear rate
$\dot{\gamma}$ (see Figure~2 of ref.~\cite{Dangelo2007})
is well described by the Carreau function:
\begin{equation}\label{powervisc}
\mu = \frac{1}{(1+
(\dot{\gamma}/\dot{\gamma_0})^2)^{\frac{1-n}{2}}} (\mu_0 -
\mu_{\infty}) + \mu_{\infty}.
\end{equation}
with $n=0.26$, $\dot{\gamma_0}=0.026 \pm 0.004\ \mathrm{s^{-1}}$, ${\mu}_0
=4500 \pm 340\ \mathrm{mPa.s}$. A shear thinning domain of shear rates in which $\mu \propto
{\dot{\gamma}}^{(n-1)}$ is bounded by two Newtonian domains. At low
shear rates ($\dot{\gamma} \ll \dot{\gamma_0}$), the
viscosity is constant with $\mu=\mu_0$. At very high shear rates
($\dot{\gamma} \gg \dot{\gamma}_\infty=\dot{\gamma_0}(\mu_0/\mu_\infty)^{1/1-n}$),
the effective viscosity should reach the  limiting value $\mu_\infty$.
Practically, and following theoretical expectations, $\mu_\infty$ is taken equal to the viscosity of the
solvent ({\it i.e.} $\mu_\infty = 1 \ \mathrm{mPa.s}$ for water) because it cannot be measured directly
(the corresponding  transition shear rate
$\dot{\gamma}_\infty \simeq 2300 \ \mathrm{s^{-1}}$ is above the useful range of
our rheological measurements ($\dot{\gamma} < 100 \ \mathrm{s^{-1}}$)).
\subsection{Characteristic velocities and shear rates in the experiments}
\label{florate}
In the present experiments, the mean  flow velocity
ranges from $U = 50$ to $400\ \mathrm{mm/s}$.
For water,  the corresponding Reynolds number defined as $Re=U \bar{a} \rho / \mu$ varies from
 $40$ and $320$: substantial
inertial effects are therefore expected~\cite{Zimmerman2004}, and will be larger  for rough fractures.
An alternative value of the Reynolds number is obtained by using  the diameter
of the fiber and its relative velocity with respect to the fluid; it remains however  higher than $1$,
confirming the need to take into account inertial effects.

Regarding the shear rate $\dot{\gamma}$,  it is equal to  zero in the center
 part of the aperture and reaches a maximum $\dot{\gamma}_w$ at the walls. For a Newtonian fluid, $\dot{\gamma}_w =6 U/a$ (neglecting the  influence of the  fiber and of  the  roughness of the walls). For shear thinning fluids~\cite{Gabbanelli2005},
 $\dot{\gamma}_w$ depends on the relative values of the shear-rate $\dot{\gamma}$
and of the parameters $\dot{\gamma_0}$ and $\dot{\gamma}_\infty$ (see Sec.~\ref{fluids}) in the flow.
At very low mean flow velocities $U$ for which $\dot{\gamma}_w \le \dot{\gamma_0}$, the viscosity would be equal
to  $\mu_0$ in the full fluid volume and the results would be the same as for a Newtonian fluid. Using
 the numerical values from Sec.~\ref{fluids}, this would require that $U < a \dot{\gamma_0}/6 = 2.16\times 10^{-3}\, \mathrm{mm/s}$,
 a value well below the present experimental range ($U > 50 \, \mathrm{mm/s}$). The fluid has therefore a shear thinning
 behavior in a part of the flow volume.

 In the opposite high velocity limit, a second Newtonian layer with  $\mu = \mu_\infty$ develops at the walls
if $\dot{\gamma}_w > \dot{\gamma}_\infty$.
Assuming a truncated power law rheological curve  (see~\cite{Gabbanelli2005}),
 the corresponding threshold value of $U$ may be estimated as:
$U = \dot{\gamma}_\infty a \left(\frac{1}{6} \right)^{1/n} \left[\frac{3n}{2n+1}\right]$. Using the numerical
values from Sec.~\ref{fluids}, this leads to $U = 2\, \mathrm{mm/s}$, again below the
experimental range of velocities used in the present work. A second  Newtonian
layer of viscosity $\mu_\infty$ appears therefore on the walls in our experiments and the effective local
viscosity increases from  $\mu_\infty$ at the walls to $\mu_0$ in the center of the gap.
The thickness $\delta z$ of the Newtonian layer of viscosity $\mu_\infty$ on the walls at the
highest experimental flow velocity $U = 400\, \mathrm{mm/s}$ may be estimated from
the relation $2\delta z/a = 1 - a \dot{\gamma}_\infty/(6U)$ valid if
$\dot{\gamma}_w \gg \dot{\gamma}_\infty$: the combined thicknesses of these two Newtonian layers
on both walls are of the order of $52\%$ of the gap $a$.
\section{Fiber transport by water in smooth and rough model fractures}
\label{sec:res}
\subsection{Fracture with smooth walls (model $F1$)} \label{planefracF1}
\subsubsection{Transport of fiber segments}
\begin{figure}[h!]\noindent\includegraphics[width=\W]{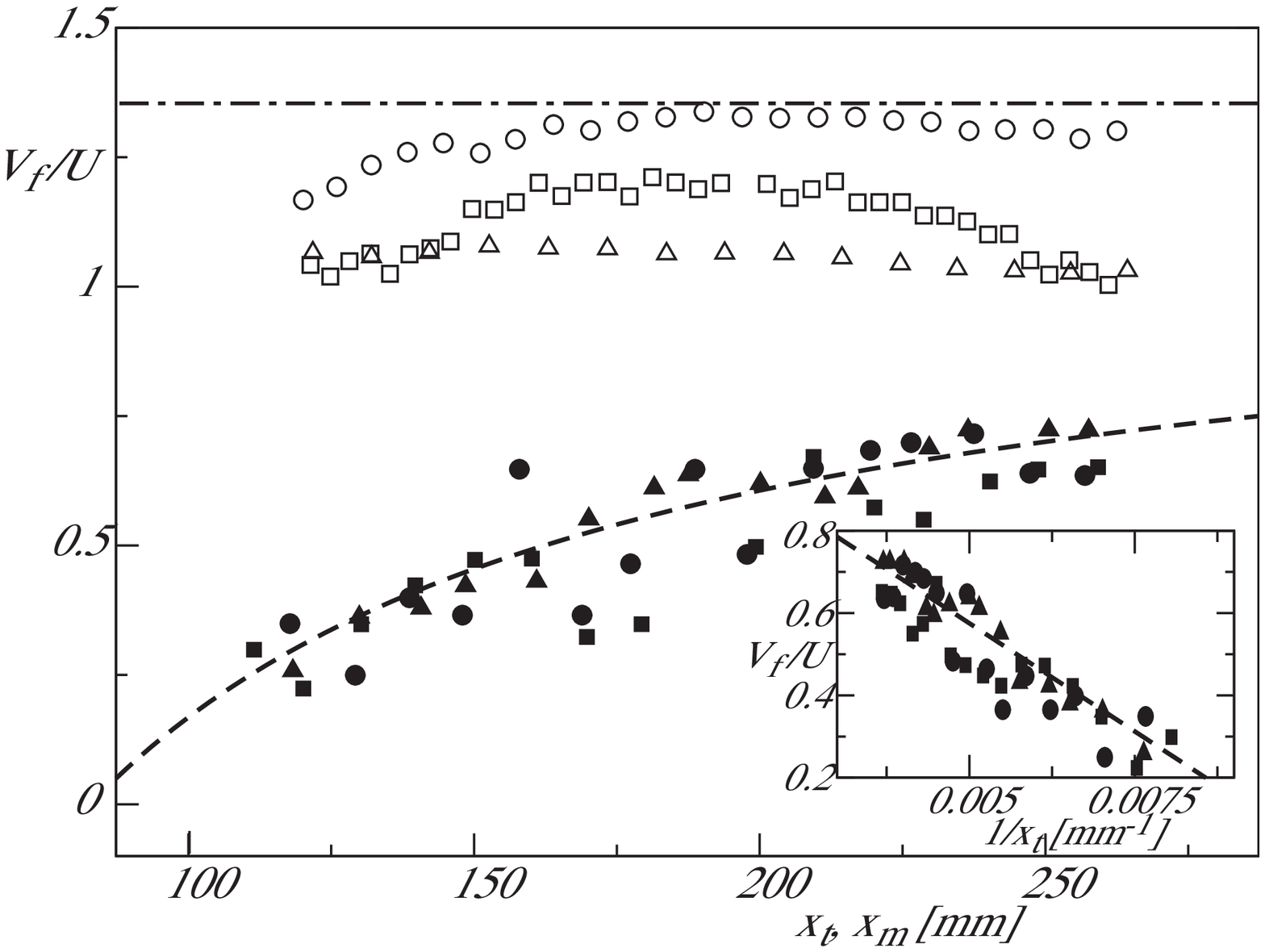}
\caption{Variation of the normalized velocities $V_f/U$ of the tip
of continuous fibers  as a function of the distance $x_t$ of the tip from the injection side of
the model:
($\blacksquare$)  $U =\ 100 \, \mathrm{mm/s}$, ($\bullet$)   $U =\,150 \ \mathrm{mm/s}$,
($\blacktriangle$)  $U =\ 290 \, \mathrm{mm/s}$.
Variation of the velocity $dx_m/dt$ of the center of mass of
fiber segments of length $\ell = 20 \mathrm{mm}$ as a function of $x_m$:
($\square$)  $U =\, 100 \, \mathrm{mm/s}$, ($\circ$)  $U =\, 150 \, \mathrm{mm/s}$,
($\triangle$)  $U =\, 290 \, \mathrm{mm/s}$.
Dashed line : $V_f/U=1.1 - 105/x_t$. Inset: variation of $V_f/U$ as a function of $1/x_t$ for
the continuous fiber (same symbols as in main graph). Dash-dotted line: estimated value
$V_f/U = 1.35$ from ref.~\cite{Sugihara-Seki1993}} \label{fig:Vplanefrac}
\end{figure}
As a first step, reference experiments were carried out on the model fracture $F1$
with flat smooth walls, starting with fiber segments of length $\ell$ ranging
from $20$ to $150\, \mathrm{mm}$ and at mean velocities $U$ of the fluid
across the aperture: $50 \le U \le 400\, \mathrm{mm/s}$. It should be noted, in addition,
 that few experimental and even numerical results (outside reference~\cite{Sugihara-Seki1993})
 are available on the transport of fibers, even in the simple geometry of parallel plane walls).
The present study is centered on the transport of fibers with their
mean axis parallel on the average to the direction of the flow within
$\pm 10^\circ$. For each set of values of $\ell$ and $U$, more than
$20$ transport experiments were realized.
In all cases, no deformation of the fibers is visible during the motion
and they reach a  velocity which is constant with the distance $x_m$
of their center of mass to the injection side within $\pm 10 \%$.
This latter result is visible in Figure~\ref{fig:Vplanefrac} in which the ratio
of the velocity of the fiber ($V_f$) and of the mean fluid velocity
 ($U$) is plotted
as a function of $x_m$ for $3$ different $U$ values (open symbols)
in a range of $1$ to $3$.
Taking into account the intrinsic dispersion
due to the injection process and the variability of the shape of the fiber,
the variations of $V_f/U$ ($\le 25 \%$) are too small to suggest any
definite dependence of $V_f/U$ on $U$.
More generally, in other experiments  performed at different velocities
and for fibers of different lengths,
 $V_f/U$ always ranges between $1$ and $1.35$ at all values of $x_m$.

These observations can be compared to the numerical simulations by
Sugihara-Seki~\cite{Sugihara-Seki1993} of the longitudinal velocity of
a neutrally buoyant circular cylinder located midway between the walls of a
plane channel. For a ratio of the diameter of the cylinder to the distance between the walls
equal to $0.5$ as in the present experiments, the normalized predicted velocity $V_c/U$
is equal to 1.35~\cite{Sugihara-Seki1993}; this is consistent with the upper
bound of the present experimental results  (dash-dotted line in Fig.~\ref{fig:Vplanefrac}).
The lower values would then correspond to cases in which the fiber is not
in the center of the interval or in which it has a wavy shape which obstructs
partly the flow field.

\begin{figure}[h!]
\noindent\includegraphics[width=\W]{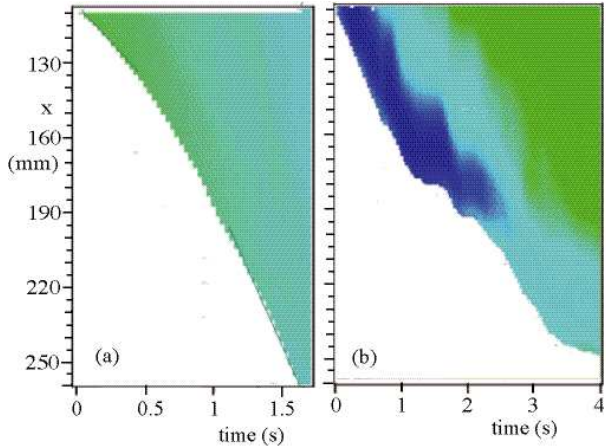}
\caption{Spatiotemporal diagram of  the  transport of a continuous fiber by a downward flow of water
 in two model fractures.
Horizontal axis: time - vertical axis: distances in the vertical
direction counted from the top of the model
(see Fig.~\ref{fig:setup}).
Color code:  transverse coordinate $y(x)$  of fiber
at the corresponding distance $x$. (a) Fracture $F1$ with smooth walls.
 ($U = 120\, \mathrm{mm/s}$). $y$ ranges from $52.9\, \mathrm{mm}$ (green shade) to
 $63.3\, \mathrm{mm}$ (light blue) - (b) Fracture $F2$ with  rough walls
 ($U = 160\, \mathrm{mm/s}$). $y$ ranges from $58.4\, \mathrm{mm}$ (green) to
 $71.5\, \mathrm{mm}$ (dark blue).
End of fiber inserted initially in the high permeability region of the fracture
($a > 1\, \mathrm{mm}$).}  \label{fig:spatio}
\end{figure}
\subsubsection{Transport of continuous fibers}
The above results are now compared to the case of a continuous fiber lying
initially at the top of the model with no external force applied and dragged
by the flowing fluid into the space between the fracture walls once its
tip has been inserted into it.
 Figure~\ref{fig:spatio}a displays a
spatiotemporal diagram obtained for such an experiment by the technique
described in Sec.~\ref{setup}. At a given time $t$,
the color is constant along $x$ in the colored zone, indicating that the fiber remains straight and
vertical. The color varies slightly with time which reflects a small global transverse displacement.
The boundary of
the colored zone marks the coordinate $x_t$ of the tip at the corresponding time $t$: its slope (proportional to the velocity $V_f$) increases with $t$.
Quantitatively, Fig.~\ref{fig:Vplanefrac} shows (dark symbols) that the normalized velocity $V_f/U$
of the tip of the fiber increases with $x_t$ ($U$ is the  mean flow velocity
in the constant aperture part of the model).

We  show now by a simple model that this variation of  $V_f/U$ reflects the lower flow velocity
 in the upper $52\, \mathrm{mm}$ of the fracture  due to the larger cross section of the flow
 (this upper part has a  ``$Y$'' shape with an aperture increasing towards the top).
We assume that  the fiber: a) remains straight and vertical so that  the velocity of all  its points is equal to $V_f$
and: b) is located in the center of the gap where the velocity has a maximum value $V_M(x)$ (assumed
to be independent of the transverse coordinate $y$).

As a very first approximation,  the local vertical viscous force $f_v$
per unit length on the fiber is taken equal to  $f_v = \mu C (V_M(x) - V_f)$ in
 which $C$ is a geometrical constant.
Due to the low angle and the small distance between the walls for $x > 0$, the flow field is
quasi parallel with a Poiseuille-like profile between the walls at a given vertical coordinate $x > 0$
(for a Newtonian fluid and neglecting flow perturbation by the fiber). In the upper fluid bath
 ($x < 0$), this result is not satisfied but the fluid velocity is low enough so that its
contribution is negligible.

In order to comply with   volume flow rate conservation, the mean fluid velocity
$\overline{V}(x)$ at the distance $x$ where the local aperture is $a(x)$ must satisfy:
$\overline{V}(x)\, a(x) = U a = cst(x)$ in which $a$ is the  the aperture in the parallel part of the cell.
where the mean velocity is $U$. The total viscous force on the continuous fiber between its lower tip at
 the vertical coordinate $x_t$ and the surface of the upper bath is then :
\begin{equation}\label{Frviscin}
 F_v(x_t) = \int_{-l_b}^{x_t} f_v(x) dx =  \mu C \left [ B U a \int_{0}^{x_t} \frac{dx}{a(x)} - (l_b+ x_t)V_f \right ]
 \end{equation}
$a(x) = a$ for $x \ge l_e$ and $a(x)$ decreases linearly from
$a_e$ to $a$ with $x$ for $0 \le x \le l_e$. As mentioned above, the velocity of the fluid
 in the upper bath ($x < 0$) has been neglected.
The coefficient $B$ would be equal to $3/2$ for a Poiseuille
profile with no obstruction effect. The integral may be readily computed and $V_f$ may be estimated
by taking $F_v = 0$: this is equivalent to neglecting inertial forces as well as friction on the walls which
has been found to be approximately valid  in the case of the previous segments (the fiber moved roughly
at the maximum velocity in the gap). This leads to:
\begin{equation}\label{Frvisc}
V_f (x_t + l_b) = B U \left [ x_t  - l_e + \frac{l_e a}{a_e -a}\ell n \frac{a_e}{a} \right ]
 \end{equation}
The ratio $V_f/U$ may then be rewritten as:
\begin{equation}\label{Vfx}
\frac{V_f}{U} =b^* \left [ 1 - \frac {l^*} {x_t +l_b} \right ]
 \end{equation}
in which $b^*$ and $l^*$ are two constants and $x_t + l_b$ is the vertical distance of the tip of the fiber to the  free surface of the bath.  Using the parameter values given in
Fig.~\ref{fig:setup} leads to the theoretical prediction:
$b^* = B = 3/2$ and $l^* =  l_b  + l_e (1 - a \, \ell n(a_e/a) /(a_e - a)) \simeq 57\, \mathrm{mm}$.
Experimentally, $b^*$ and $l^*$  can be determined conveniently by plotting
$V_f/U$ as function of $1/(x + l_b)$ (inset of Fig.~\ref{fig:Vplanefrac}).
The variation is linear as expected and a  linear regression
over the data gives  the values  $b^*=1.1 \pm 0.1$ and $l^*=105 \pm 15 \, \mathrm{mm}$.
These values are valid for all three velocities $U$ investigated and also lead,
as might be expected,  to a good fit (dotted line) in the main graph of
of Fig.~\ref{fig:Vplanefrac}.
The difference between the experimental and theoretical coefficients likely reflects
perturbations of the flow profile by the fiber so that both $B$ and $C$ may differ
from the theoretical value and depend on $x$.
 Finally, neglecting the influence of the inlet would only be possible for  $x >> l^*$.
 This is not the case here since the global fracture length is only
$\simeq 3 l^*$.
\subsection{Fiber transport in a wedged  fracture with rough walls}
\label{sec:F2W}
\begin{figure}
\noindent\includegraphics[width=\W]{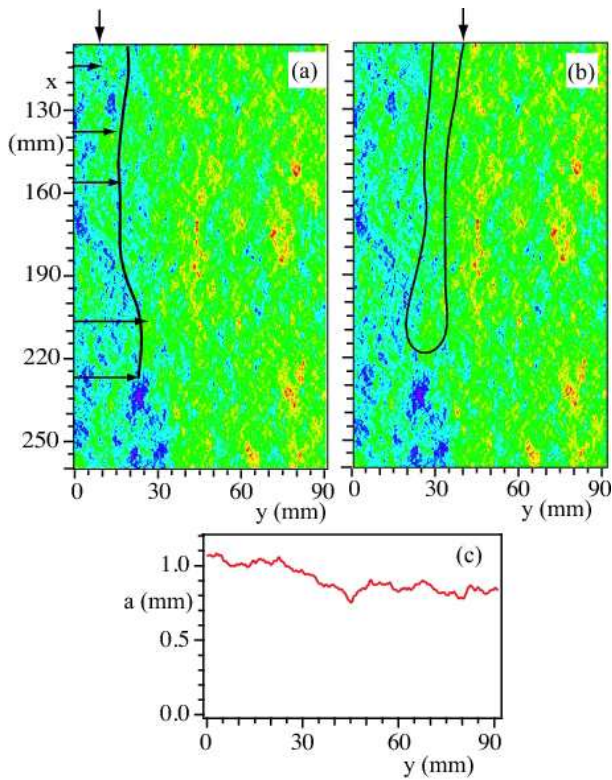}
\caption{(a-b)Snapshots of continuous fibers (in black) during experiments realized in model $F2$ with two transverse
injection distances (vertical arrows). The images of the fiber are superimposed onto
a color coded map of the aperture field (red: $a \le \, 0.4\, \mathrm{mm}$,
blue: $a \ge 1.4 \, \mathrm{mm}$). Field of view: $90 \times 150\, \mathrm{mm}$.
Distances in the vertical $x$ direction are counted from the top of the model
(see Fig.~\ref{fig:setup}). Vertical arrows mark the location of the entry point of
the fibers inside the field of view.
a)  $U = 160 \, \mathrm{mm/s}$; horizontal arrows: $5$ pinning positions
b)  $U = 250 \, \mathrm{mm/s}$. The fiber is pinned outside the image at
a point of coordinates ($x \,=\, 150 \mathrm{mm}$ and $y \,=\, 60 \mathrm{mm}$).
 (c)  variation
of the aperture averaged along the distance $x$ parallel to the flow as
a function of the transverse distance $y$. }
\label{fig:fiberwedge}
\end{figure}
While, for fractures with smooth plane walls, the fluid velocity is constant with $x$ and $y$
 and no deformation
of the fibers is observed, the flow  field is very complex in rough
fractures with large shear stresses in the fracture plane. If the fiber is sufficiently flexible,
it may get bent by these stresses. Also, friction of the fibers with the walls is larger
than for the smooth fracture of Sec.~\ref{planefracF1} and it screens  the influence of the
 inlet much more efficiently.
All the experiments with rough walls described in the rest of the paper have
been realized with a continuous fiber.

Fiber transport by a water flow has been studied in fracture $F2$ which
has a self-affine roughness geometry with a small angle between
 the mean planes of the walls.  In this model, one has both random variations of the velocity
due to the roughness and a mean transverse gradient.
Since the mean aperture ${\bar a}(y)$ varies continuously with $y$
(Fig.~\ref{fig:fiberwedge}c), this allows
us to test the minimum value of $\bar a(y)$ (averaged over the length $L$) for which
fiber transport all along the  fracture is possible.
In the experiments using water as the carrier fluid, this minimum value
was found to be of the order of $1\,  \mathrm{mm}$ (typical mean aperture in the left side of
the maps of Figs.~\ref{fig:fiberwedge}). A view of a fiber obtained during such an experiment  is
superimposed onto the aperture field in Figure~\ref{fig:fiberwedge}a.
 In the low aperture part, most fibers get pinned in the first cm past the
 initial large aperture section at the top of the model.
 However, in a few experiments performed at high flow rates ($U \ge 250 \,  \mathrm{mm/s}$),
 the tip of the fiber gets first pinned but a loop builds up downflow
 of the tip~(Fig.~\ref{fig:fiberwedge}b);  then, the bottom of the loop moves sideways
into the high aperture region where the velocity is largest. Finally, the loop pursues its
 downward motion until it reaches the bottom of the model.

Let us now concentrate on the dynamics of a fiber injected in the
high permeability (large aperture) region at the left.
 Figure \ref{fig:spatio}b displays a spatiotemporal diagram of such an experiment:
it differs strongly from the diagram of  Fig.~\ref{fig:spatio}a obtained for a model
with plane walls. The motion of the tip is not continuous but corresponds
to a sequence of ``stop'' and ``go'' phases marked respectively by horizontal and oblique
sections of the boundary of the colored zone in Fig.~\ref{fig:spatio}b. The duration of
individual pinning events may range from less than $1 \  \mathrm{s}$ to nearly
$10 \ \mathrm{s}$ (about $10 \ \%$ of the total transit time of the fiber along the fracture).
One also observes that pinning is always initiated at the tip at the fiber and not in
the zones located behind it.

The  sites where the fiber gets pinned are shown by horizontal arrows in
 Fig. \ref{fig:fiberwedge}a.  Several (but not all) of them are located in regions where the
aperture is below $1\  \mathrm{mm}$ (yellow-green shades in Fig. \ref{fig:fiberwedge}a).
This  is still twice the diameter of the fiber so that
pinning cannot result solely from blockage in a constriction
too small to accomodate the fiber.  Moreover, the fiber is observed
to cross  other regions of the fracture with a similar aperture without
getting stopped.
Blockage of the fibers may also result from a large local slope of the surfaces, even if  the local
spacing remains large. In this case, the hydrodynamic forces may be too small
to induce enough bending of the fibers so that they follow the slope of the surface and do not
touch it. Examining the geometry of the individual wall surfaces around the
pinning points tends to confirm this hypothesis.
Finally, the inertia of the fiber may also favour
contacts with the walls and blockage: this effect should increase with
the fluid velocity.

\begin{figure}[h!]
\noindent\includegraphics[width=\W]{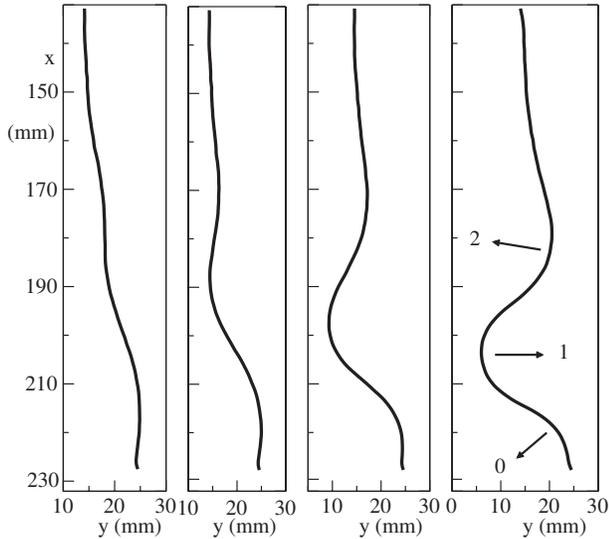}
\caption{Variation of the fiber geometry during the fifth pinning event in
Fig.~\ref{fig:fiberwedge}a (lowest horizontal
arrow). The leftmost (respectively rightmost) picture
displays the shape of the fiber immediately after pinning (respectively before depinning).
Time lapse between pictures: $0.166\ s$; total duration of trapping event: $1\ s$.
Arrows: location of points of maximum curvature referred to as $0$, $1$ and $2$.}
\label{fig:fiberpinwater}
\end{figure}
More information on the pinning-depinning process is obtained from the geometry of the fiber.
The variations of the code color with time and distance in
Fig.~\ref{fig:spatio}b indicate that, in contrast to the case of a smooth fracture
(Fig.~\ref{fig:spatio}a), the fiber does not remain straight: it displays a buckling instability
and meanders
of shape and location varying with time appear. These variations are visualized directly
in  Figure~\ref{fig:fiberpinwater} displaying the deformations of the fiber. During the pinning
event, the tip of the fiber remains motionless while three bumps appear behind it;
their amplitude increases with time while they propagate toward the tip. The deformation of the
rear part  of the fiber is much weaker while, in some cases, it gets translated sideways.

\begin{figure}[h!]
\noindent\includegraphics[width=\W]{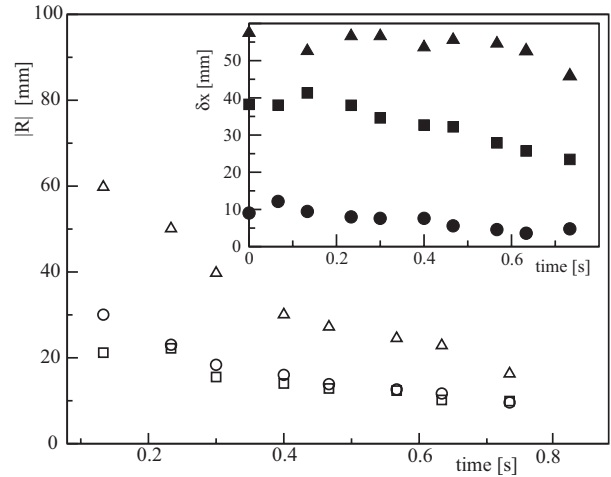}
\caption{Variation with time of the absolute values $|R|$ of the radius of curvature of the fiber
at the $3$ maximum curvature points
designated by arrows in Fig. \ref{fig:fiberpinwater}. Inset:
time variation of the vertical distances  $\delta x$ between these  $3$ points
and the fiber tip. Symbols for point $0$: ($\circ$,$\bullet$); for point $1$: ($\square$,$\blacksquare$)
and for point $2$: ($\triangle$,$\blacktriangle$).}
\label{fig:fiberpinwatercurve}
\end{figure}
We characterized quantitatively the deformation of the fibers  by
the maximum absolute value $|R|$ of the radius of  curvature in the three bumps
and by the distance $\delta x$ of the corresponding points to  the tip of the fiber
(measured along axis $x$). The variation of $|R|$ and $\delta x$ with time
for all three regions in this same experiment  is displayed in Fig.~\ref{fig:fiberpinwatercurve}.
The distance $\delta x$ decreases with time with a relative velocity of the order of
$10\, \mathrm{mm/s}$ lower than the fluid velocity ($160\, \mathrm{mm/s}$ in Fig.~\ref{fig:fiberwedge}a) and
similar for the three points of interest while the corresponding
radii $|R|$ decrease  strongly. The deformation is more important in the two
first bumps for which the radius of curvature is of the order of $9.5\  \mathrm{mm}$ at the time
when depinning  occurs.  These deformations reflect the accumulation of bending elastic energy
behind the tip which, finally, gets large enough to overcome the friction forces with the wall.
Similar scenarios were observed in all cases in which pinning takes place even though the time
scales varies.

\begin{figure}
\noindent\includegraphics[width=\W]{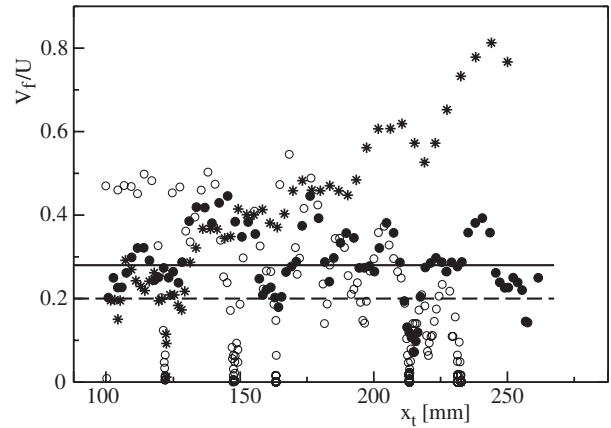}
\caption{Variation of the ratio $V_f/U$ of the fiber and fluid velocity as a
function of the distance $x_t$ from the inlet of model $F_2$ for three experiments with
continuous fibers injected at a same location. ($\circ$) and ($\bullet$): water flowing
at respective velocities  $U=160\ \mathrm{mm/s}$ and
$210\ \mathrm{mm/s}$. ($\ast$): flow of a polymer solution at  $U=220\ mm/s$.
Dotted and solid lines: average of fiber velocity over field of view respectively
 for $U = 160$ and $210\ \mathrm{mm/s}$.} \label{fig:velocitywedge}
\end{figure}
The pinning/depinning process should depend significantly on the hydrodynamic forces and,
therefore, on the flow  velocity $U$.
This influence  is analyzed in Figure~\ref{fig:velocitywedge} displaying
the variation with distance of the normalized velocity $V_f/U$ of the fiber tip for
two different values of $U$.
At the lower velocity $U=160\, \mathrm{mm/s}$, the fiber displays a stop and go motion already
visible qualitatively in
Fig.~\ref{fig:spatio} corresponding to the same experiment.
The velocity cancels out at each pinning point and increases sharply after the release of the fiber
to a value much higher than the time average (horizontal dashed line).
For $U = 210\, \mathrm{mm/s}$, the
variations with time of the fiber velocity with respect to the mean value are of smaller amplitude
 (-100\%,+50\%) instead of (-100\%,+150\%).  Also, the fiber only stops completely once while
$V_f$ decreases by less than  $40\%$ on other pinning sites: these are located at the same distances
 $x$ as for $U=160\, \mathrm{mm/s}$. These results imply that the influence on pinning of inertia forces
 dragging the fibers towards the walls is small.
 The normalized averages $\overline{V_f}/U$ of the fiber velocity with time are equal to
$0.2$ and $0.28$ respectively for $U=160$ and $210 \, \mathrm{mm/s}$. Even if the influence
of the low velocity near the pinning sites is removed from this average, $\overline{V_f}/U$ is still only
of the order of $0.3$ for both  $U$ values. This  is lower by a factor of $3$  than the values  measured
in  model fractures with smooth plane walls (see Fig.~\ref{fig:Vplanefrac}).
\section{Fiber transport by  shear thinning fluids in rough model fractures}
\label{sec:FRNN}
\subsection{Fiber transport  in fracture $F2$}
\label{sec:F2NN}
The influence of rheology has been studied specifically  by replacing  water by a shear thinning solution of characteristics
 discussed in section~\ref{fluids} in the same model $F2$ as above.
 In Fig.~\ref{fig:velocitywedge}, data obtained
with this fluid are compared to the variations for
water discussed in the previous section (the points of injection of the
fibers are the same and correspond to the high aperture path). At a similar velocity
($\simeq 220\, \mathrm{mm/s}$), the ratio $V_f/U$ is larger for the
shear thinning solution than for water, except at short  distances.
The amplitude of the velocity fluctuations is also smaller but they
still take place at the same distances $x$ as the  pinning sites observed
in low velocity water experiments. The motion of the fiber remains therefore
slightly influenced by the local asperities encountered
along its path, although much less than for water.

Globally, using a shear thinning fluid instead of water reduces significantly the
influence of the roughness of the walls on the motion of the fibers: their dynamics may be then
 much more similar to that reported in Sec.~\ref{planefracF1}  for the plane fracture $F1$.
In particular, the increase of $V_f/U$ with the distance $x$ indicates that the reduced drag
forces on the parts of the fiber located at the inlet influence the global motion of the fiber,
 like for fracture $F1$ but unlike the experiments of Sec.~\ref{sec:F2W} with water.

\begin{figure}[h!]
\noindent\includegraphics[width=\W]{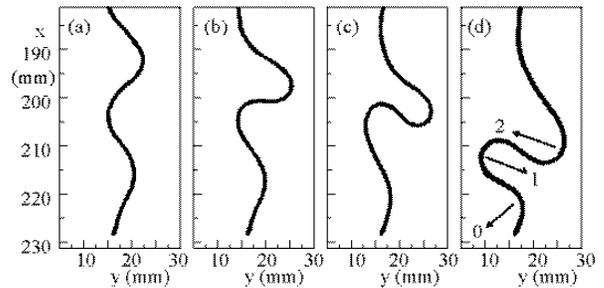}
\caption{Variation of the fiber geometry during a pinning event for a polymer flow
($U = 230\, \mathrm{mm/s}$) in the low aperture part of fracture $F2$. The leftmost (respectively rightmost) picture
displays the shape of the fiber $0.1\ \mathrm{s}$ after pinning (respectively just
before depinning). The definitions of points $0$, $1$ and $2$ are the same
as  in Fig.~\ref{fig:fiberpinwater}} \label{fig:fiberpinpol}
\end{figure}
\begin{figure}[h!]
\noindent\includegraphics[width=\W]{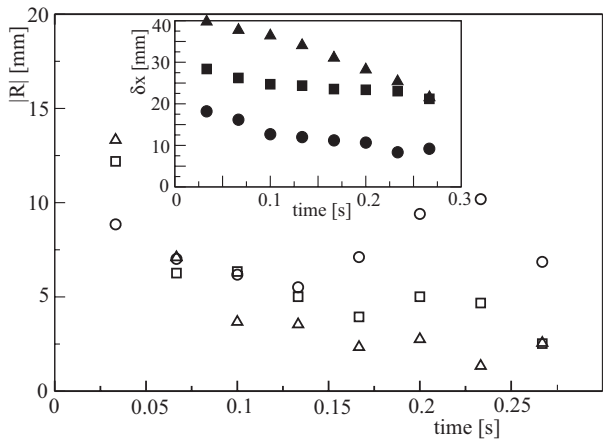}
\caption{Variation with time of the radius of the fiber at the $3$ maximum curvature points
designated by arrows in Fig. \ref{fig:fiberpinpol}(d). Inset:
time variation of the vertical distances  $\delta x$ between these  3 points
and the fiber tip. Symbols for point $0$: ($\circ$,$\bullet$); for point $1$: ($\square$,$\blacksquare$) and for point $2$: ($\triangle$,$\blacktriangle$).}
\label{fig:fiberpinpolcurve}
\end{figure}

The above results correspond to fibers moving, like in Sec.~\ref{sec:F2W},  in the high
aperture part of the width of the wedged fracture $F2$.
However, fibers may propagate even in the low aperture part, if the fluid is shear
thinning while they get blocked  or build up side loops with water (see Fig.~\ref{fig:fiberwedge}).  The
corresponding motion is then  not continuous but of a ``stop'' and ``go'' type:
the lower mean value of the aperture  in these regions increases its relative fluctuations
and there are more potential pinning sites.
As already reported for  water in Sec.~\ref{sec:F2W},
a buckling instability occurs and  fibers are deformed as they get pinned (Fig.~\ref{fig:fiberpinpol}).
  Like for water, these deformations are localized near the
tip of the fiber and move towards it as they develop; their amplitude and curvature may
however  become significantly larger than for water before depinning takes place
(see Fig.~\ref{fig:fiberpinwater} for comparison). In Fig.~\ref{fig:fiberpinpol}, for instance,
 an overhang appears  as bump $1$ is overtaken by $2$ as shown by the two upper sets
 of points in the inset.
Also, the variation of the radius of curvature with time is more complex and not always
 monotonic (Fig.~\ref{fig:fiberpinpolcurve}),
reflecting the interaction between the different loops.
The radii of curvature of loops $1$ and $2$ both become as low as
 $\simeq 2\, \mathrm{mm}$ or less: this is lower than the radius of loop $0$
and, also, much below  the radii observed with water.

These larger deformations may
result both from the larger  hydrodynamical forces in the polymer flow due to
its higher mean viscosity and from
the higher  depinning energies in the narrow part of the wedge.
The increased value of the hydrodynamical forces is also reflected in the
faster variations of the distances $\delta x$ (see insets of Fig.~\ref{fig:fiberpinwater}
and Fig.~\ref{fig:fiberpinpolcurve}).
\subsection{Fiber transport by shear thinning fluids in fracture with rough parallel walls.}
\label{F3}
\subsubsection{Aperture field and anisotropy of  fracture $F3$.}
\label{AFF3}
The results of the previous section demonstrate that, using a polymer solution, it is
possible to transport fibers along the full length of model $F2$,
even in the lower mean aperture regions. The ``wedge like''  geometry of this model has allowed us to estimate
the influence of the mean aperture $\bar{a}$ by varying the distance $y$ at the inlet of the model
where the fiber is injected: however, this geometry may have unwanted effects
like the sideways motions displayed in Fig.~\ref{fig:fiberwedge} and it may mask the influence of
local aperture fluctuations.
The present section deals therefore with experiments realized with another model ($F3$) using
model rough walls with a similar geometry as $F2$ but with
their mean planes  parallel (see Figs.~\ref{fig:fiberparallel} for the corresponding aperture
map and transverse mean aperture profile).
As mentioned above, the fracture walls are two perfectly matching  self-affine surfaces
which are first put in contact and then pulled  apart  before introducing a small
 relative shear displacement.

As shown by several authors~\cite{NAS,Gentier1997,Yeo1998}, this geometry results in
an anisotropy of the aperture field and of the permeability: flow is easier in the direction normal to
 the shear displacement.
In the present experiments, the mean flow ($x$ direction) is perpendicular to the  shear ($y$ direction)
and should therefore be oriented in the direction of highest permeability.
The distribution of the color patches in the maps Fig.~\ref{fig:fiberparallel}a-b indicates qualitatively
that the distance  over which deviations of the aperture from its mean value persist is larger along
 $x$ than along $y$.  For instance, a long continuous vertical string of blue patches (corresponding
 to a high local aperture) is visible and is associated with a maximum in the transverse profile of
Fig.~\ref{fig:fiberparallel}c.
This suggests a channelization effect related to the anisotropy of the permeability.
Recent work has indeed shown~\cite{Auradou2006,Auradou2008} that, in this
configuration,  the flow field could be described approximately as
a set of channels of constant hydraulic aperture and  parallel to the direction $x$:
this allowed, for instance, to predict the overall geometry of the displacement fronts
between two miscible fluids.
We shall analyze now how this feature influences the transport of fibers in model $F3$.
\subsubsection{Fiber transport in model $F3$.}
\label{FTF3}
\begin{figure}[h!]
\noindent\includegraphics[width=\W]{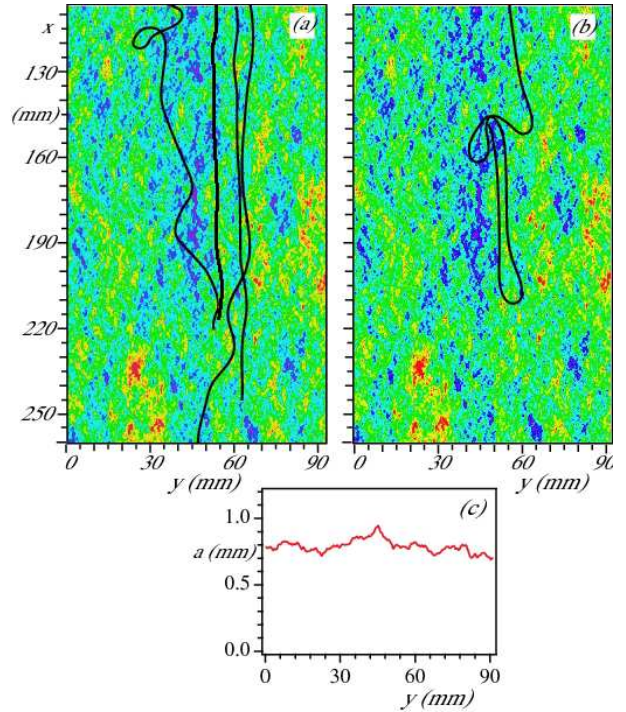}
\caption{(a) Snapshots of $4$ fibers injected into a polymer flow in fracture $F3$ at different distances
overlaid onto a map of the aperture field (mean fluid velocity $U = 215\, \mathrm{mm/s}$).
(b) Snapshot of a fiber coiled up after getting pinned inside
the fracture ($U = 242\, \mathrm{mm/s}$). A time sequence of the coiling process is displayed in Fig.~\ref{fig:fiberpinknot}.
The color code corresponds to the aperture (red: $a \le \, 0.5 \, \mathrm{mm}$, blue:
$a \ge \, 1.25 \, \mathrm{mm}$).
Field of view: $90 \times 150\, \mathrm{mm}$. Distances in the vertical $x$ direction are counted from the top of the model
(see Fig.~\ref{fig:setup}).
(c): variations of the aperture averaged along the distance $x$ parallel to the flow as
a function of the transverse distance $y$.} \label{fig:fiberparallel}
\end{figure}
In Figure~\ref{fig:fiberparallel}a, snapshots of four fibers injected at different
transverse distances $y$ are overlaid onto the aperture field of model $F3$.
Three of them are weakly deformed and kept moving afterwards down to the lower end of the model.
One notices that their location does not coincide exactly with the zone of highest aperture
of the model although it is not far from it.
The fourth fiber (at the left on the image) got pinned as
it moved out of a high aperture region into a less open one;  the
shape of this fiber is also more strongly distorted, possibly reflecting a larger velocity gradient
in the region where the fiber moves.
In model $F3$, therefore, fiber transport is not solely determined by the distribution of the aperture
even though it has a strong influence: other parameters like the local gradients of the aperture and
slope of the surface or, yet,  inertia effects may also be of importance. Some of the observations
on fracture $F2$  previously suggested similar results.
\subsubsection{Irreversible pinning and fiber entanglement in model $F3$.}
\label{PEF3}
\begin{figure}[h!]
\noindent\includegraphics[width=\W]{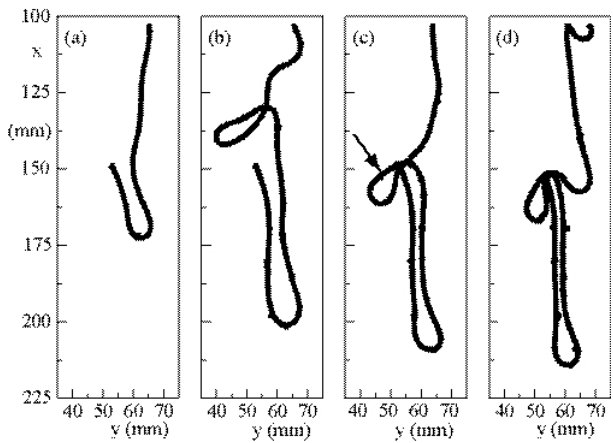}
\caption{Evolution of the fiber as function of time during the
build-up of a coil for the flow of a polymer solution in  fracture $F3$ (see Fig.~\ref{fig:fiberparallel}b).
>From left to right, the time elapsed after pinning is respectively  $0.63$, $1.13$, $1.3$ and $1.8\
s$.} \label{fig:fiberpinknot}
\end{figure}
Up to now, we have discussed mostly pinning events of finite duration like in  Figs.~\ref{fig:fiberpinwater}
and \ref{fig:fiberpinpol}. Irreversible pinning has however also been observed, for instance
 in the narrow parts
of fracture $F3$; this  often leads to an entanglement
of the fiber around the pinning site and to the build up of intertwined loops like
 in Figs.~\ref{fig:fiberpinknot} and \ref{fig:fiberparallel}b.
Let us discuss in more detail the process leading to the configuration displayed
 in Fig.~\ref{fig:fiberparallel}b following pinning in a narrower zone (yellow shade region
 near the top of the coil).
At first, the fiber gets deformed  like in the case of transient pinning  (see
Sections~\ref{sec:F2W} and  \ref{sec:F2NN} and Figs.~\ref{fig:fiberpinwater}
and \ref{fig:fiberpinpol}).
Then, while the tip remains pinned,  bump $1$ overtakes it so that a loop appears
at the right of the pinning site and move downwards
(Fig.~\ref{fig:fiberpinknot}a). At the same time other parts of the fiber located
 farther upstream move leftwards and build up a second, closed
 loop (arrow in Fig.~\ref{fig:fiberpinknot}). A key point is the fact that the friction forces
with the walls in the parts of the fiber located close to the pinning site are large enough to
prevent hydrodynamic forces on fast moving parts of the fiber from dragging downward
the rest of its length. Tension  forces on the upstream parts are then quite low so that
loops and deformations can easily appear. Finally, both loops move down: they get
locked on the pinning site and dangle on both sides  (Fig.~\ref{fig:fiberpinknot}c).
This further reduces the tension on the upstream fiber sections so that new loops
start to build up (Fig.~\ref{fig:fiberpinknot}d).
\section{Conclusion}
In the present work, several clear-cut results have been obtained experimentally
on the conditions for achieving the transport of flexible fibers in fractures.
First, while in smooth fractures the friction of the fibers with the walls is negligible and does
not influence their motion, its influence has been verified to be much stronger for rough walls:
this  may result in pinning and blockage or, at least, in a jerky  progression instead of a
continuous one.

Next, the aperture of the fracture (and particularly its ratio to the fiber diameter) is
also a key factor, but with a more subtle influence  than expected.
On the one hand,  in the wedge-shaped rough model $F2$ with  a global transverse
gradient of the mean aperture, fiber transport is, as expected, easier in the more
open parts of the wedge  than in narrower ones
(however pinning may occur in open regions).
In the rough fracture $F3$ with globally  parallel walls and a channelized flow field, fiber
motion is  also easier in high permeability regions.
On the other hand, the trajectory of the fibers does not coincide fully with the paths of highest
aperture  while pinning is also not always localized  at minimal aperture points.
Other factors such as the local slope of the walls and inertia forces  pushing the fibers
 towards the solid surfaces must therefore play  an important part.
In the present experiments, the mobility of the fibers appears to increase with the mean flow velocity:
further experiments will however be necessary to determine whether, at high velocities,
some fibers may be forced into narrow zones by inertia effets and get blocked there.

In order to evaluate the transport of fibers in natural or
industrial fracture systems of practical interest, different configurations of the aperture
field need to be studied such as,  for
instance, a channelization perpendicular to the mean flow.
Using our experimental set up,  this may be investigated
by changing the direction of the shear displacement between the two complementary
rough walls with respect  to the flow (more precisely by comparing for two identical
complementary rough walls results obtained when the shear displacement $\vec{u}$ is
parallel and perpendicular to the mean velocity $\vec{U}$).
 The value of Hurst's exponent $H$ for the fracture walls may also influence fiber transport
by determining the relative height of rugosities of different sizes in the plane of the fracture: it
will be informative to compare the present measurements corresponding to $H = 0.8$ to others achieved
for  $H = 0.5$ (a value encountered in such materials as sandstone and sintered glass).

A third key observation is the enhancement of the mobility of the fibers when water is replaced
by a shear thinning fluid at a same flow velocity. For instance, fiber propagation along the full length of
rough fractures with a mean aperture  of less than twice the  fiber diameter becomes
possible. Fiber motion is observed (although in a  ``stop and go''  manner)
in zones  where they got stuck in water flows while, in regions of high aperture the displacement
is faster and more continuous. Also, the influence of the interaction of the fibers with the walls is much
reduced and the minimum value of the ratio of the mean aperture to the diameter of the fiber
for observing transport is lower. This increased  mobility
in shear thinning flows may result from larger hydrodynamic forces (the effective viscosity
of the polymer solutions  is generally higher than that of water).
Other possible explanations must however be investigated : for instance, the  fiber is
 more frequently localized in the center of the gap between the walls than in Newtonian fluids.
Also, at high velocities, a layer of low viscosity Newtonian fluid appears near the walls and may
influence the transport of the fibers.
In order to understand this effect, it will be also important to compare the present results
with those obtained with Newtonian fluids of different viscosities and other  non Newtonian fluids
such  as shear-thinning fluids with different characteristic exponents and  yield stress or viscoelastic fluids.

The transport of fibers is also strongly influenced by their deformation under the effect of
hydrodynamic forces which is particularly important during pinning events.
In the case of intermittent pinning marked by a jerky motion, the amplitude of the deformation
of the fiber increases with time and the elastic energy accumulates until the front   tip is released
and the fiber recovers roughly its initial shape. The amplitude of the deformations and the curvatures
observed were larger for experiments with the polymer solution than for water.
In the case of permanent pinning, the amplitude of
the deformations may increase further: loops appear and  pile up on the pinning site, finally leading
to entanglement. These results show qualitatively that, even for very flexible fibers, the influence
of the elastic modulus on their transport by the flow cannot be neglected.

In future work, systematic investigations will be needed to characterize more quantitatively
 the dependence of the pinning and entanglement processes on the control parameters of
 the experiments. A first important question is the dependence of the motion of the fibers on their flexibility
 and on the value of their diameters compared both to the mean fracture aperture $\overline{a}$ and
 to its standard deviation $\sigma_a$.
Pinning effects also depend clearly on the interaction of the fibers and the walls: this involves
friction between them as well as well as blockage of the fiber motion by  the wall roughness: their relative
magnitude will need to be investigated.
Overall, as shown for instance by Fig.~\ref{fig:velocitywedge}

Finally, while the present experiments greatly help to determine the conditions and characteristics for fiber
transport in single fractures,  further work will be needed to understand
their transfer from one fracture to another in a fracture network.\\
\begin{acknowledgments}
We are indebted to R. Pidoux for the realization of
the experimental setup. We are grateful to B. Semin, E.J. Hinch and J.
Koplik for their enlightening comments. This work was supported by the Schlumberger
Foundation in Paris and by  Schlumberger-Doll Research in Boston through a
fellowship and a research grant. We also benefited from
funding by  by CNRS and ANDRA through the EHDRA (European Hot Dry
Rock Association) and PNRH programs.
\end{acknowledgments}

\begin{thebibliography}{31}
\expandafter\ifx\csname
natexlab\endcsname\relax\def\natexlab#1{#1}\fi
\expandafter\ifx\csname bibnamefont\endcsname\relax
  \def\bibnamefont#1{#1}\fi
\expandafter\ifx\csname bibfnamefont\endcsname\relax
  \def\bibfnamefont#1{#1}\fi
\expandafter\ifx\csname citenamefont\endcsname\relax
  \def\citenamefont#1{#1}\fi
\expandafter\ifx\csname url\endcsname\relax
  \def\url#1{\texttt{#1}}\fi
\expandafter\ifx\csname urlprefix\endcsname\relax\def\urlprefix{URL
}\fi \providecommand{\bibinfo}[2]{#2}
\providecommand{\eprint}[2][]{\url{#2}}

\bibitem[{\citenamefont{Stockie and Green}(1998)}]{Stockie1998}
\bibinfo{author}{\bibfnamefont{J.}~\bibnamefont{Stockie}} \bibnamefont{and}
  \bibinfo{author}{\bibfnamefont{S.}~\bibnamefont{Green}},
   {\bibinfo{title} {\bibnamefont{``Simulating the motion of flexible pulp fibres
   using the immersed boundary method''}}}, \bibinfo{journal}{J.
  of Comput. Phys.} \textbf{\bibinfo{volume}{147}}, \bibinfo{pages}{147--165}
  (\bibinfo{year}{1998}).

\bibitem[{\citenamefont{Yasuda et~al.}(2002)\citenamefont{Yasuda, Mori, and
  Nakamura}}]{Yasuda2002}
\bibinfo{author}{\bibfnamefont{K.}~\bibnamefont{Yasuda}},
  \bibinfo{author}{\bibfnamefont{N.}~\bibnamefont{Mori}}, \bibnamefont{and}
  \bibinfo{author}{\bibfnamefont{K.}~\bibnamefont{Nakamura}},
 {\bibinfo{title} {\bibnamefont{``A new visualization technique for short fibers in a slit
  flow of fiber suspensions''}}},
  \bibinfo{journal}{International Journal of Engineering Science}
  \textbf{\bibinfo{volume}{40}}, \bibinfo{pages}{1037--1052} (\bibinfo{year}{2002}).

\bibitem[{\citenamefont{Lagomarsino et~al.}(2005)\citenamefont{Lagomarsino,
  Pagonabarraga, and Lowe}}]{Lagomarsino2005}
\bibinfo{author}{\bibfnamefont{M.}~\bibnamefont{Lagomarsino}},
  \bibinfo{author}{\bibfnamefont{I.}~\bibnamefont{Pagonabarraga}},
  \bibnamefont{and} \bibinfo{author}{\bibfnamefont{C.}~\bibnamefont{Lowe}},
{\bibinfo{title} {\bibnamefont{``Hydrodynamic induced deformation
and orientation of a microscopic elastic filament.''}}},
  \bibinfo{journal}{Phys. Rev. Lett.} \textbf{\bibinfo{volume}{94}},
  \bibinfo{pages}{148104} (\bibinfo{year}{2005}).

\bibitem[{\citenamefont{Lowe}(2003)}]{Lowe2003}
\bibinfo{author}{\bibfnamefont{C.}~\bibnamefont{Lowe}},
 {\bibinfo{title} {\bibnamefont{``Dynamics of filaments: modelling the
  dynamics of driven microfilaments''}}},
  \bibinfo{journal}{Philos. Trans. R. Soc. London B}
  \textbf{\bibinfo{volume}{358}}, \bibinfo{pages}{1543--1550} (\bibinfo{year}{2003}).

\bibitem[{\citenamefont{Purcell}(1977)}]{Purcell1977}
\bibinfo{author}{\bibfnamefont{M.}~\bibnamefont{Purcell}},
 {\bibinfo{title} {\bibnamefont{``Life at low Reynolds number''}}},
\bibinfo{journal}{Am. J. Phys.} \textbf{\bibinfo{volume}{45}},
  \bibinfo{pages}{3--11} (\bibinfo{year}{1977}).

\bibitem[{\citenamefont{Selker et~al}(2006)}]{Selker2006}
\bibinfo{author}{\bibfnamefont{J.}~\bibnamefont{Selker}},
  \bibinfo{author}{\bibfnamefont{M.}~\bibnamefont{van de Giesen}},
   \bibinfo{author}{\bibfnamefont{N.}~\bibnamefont{Westhoff}},
 \bibinfo{author}{\bibfnamefont{W.}~\bibnamefont{Luxemburg}}
 and  \bibinfo{author}{\bibfnamefont{M.B.}~\bibnamefont{Parlange}},
{\bibinfo{title} {\bibnamefont{``Fiber optics opens window on stream
dynamics''}}},
  \bibinfo{journal}{Geophys. Res. Lett.}
  \textbf{\bibinfo{volume}{33}}, \bibinfo{pages}{L24401} (\bibinfo{year}{2006}).

\bibitem[{\citenamefont{on~Fracture~Characterization and Flow}(1996)}]{NAS}
\bibinfo{author}{\bibfnamefont{National Committee}
  \bibnamefont{on~Fracture~Characterization}} \bibnamefont{and}
  \bibinfo{author}{\bibfnamefont{Ffluid}~\bibnamefont{Flow}},
  \emph{\bibinfo{title}{Rock Fractures and Fluid Flow: Contemporary
  Understanding and Applications}} (\bibinfo{publisher}{National Academy
  Press}, \bibinfo{address}{Washington, D.C.}, \bibinfo{year}{1996}).

  \bibitem[{\citenamefont{Adler and Thovert}(1999)}]{Adler1999}
\bibinfo{author}{\bibfnamefont{P.}~\bibnamefont{Adler}} \bibnamefont{and}
  \bibinfo{author}{\bibfnamefont{J.-F.} \bibnamefont{Thovert}},
  \emph{\bibinfo{title}{Fractures and Fracture Networks}}
  (\bibinfo{publisher}{Kluwer Academic Publishers},
  \bibinfo{address}{Dordrecht, The Netherlands}), (\bibinfo{year}{1999}).

\bibitem[{\citenamefont{Sugihara-Seki}(1993)}]{Sugihara-Seki1993}
\bibinfo{author}{\bibfnamefont{M.}~\bibnamefont{Sugihara-Seki}},
 {\bibinfo{title} {\bibnamefont{``The motion of an elliptical cylinder in channel flow at low Reynolds numbers''}}},
  \bibinfo{journal}{J. Fluid. Mech.} \textbf{\bibinfo{volume}{257}},
  \bibinfo{pages}{575--596} (\bibinfo{year}{1993}).

\bibitem[{\citenamefont{Petrich and Koch}(1998)}]{Petrich1998}
\bibinfo{author}{\bibfnamefont{M.}~\bibnamefont{Petrich}} \bibnamefont{and}
  \bibinfo{author}{\bibfnamefont{D.}~\bibnamefont{Koch}},
  {\bibinfo{title} {\bibnamefont{``Interactions Between Contacting Fibers''}}},
  \bibinfo{journal}{Phys. Fluids} \textbf{\bibinfo{volume}{10}},
  \bibinfo{pages}{2111-2113} (\bibinfo{year}{1998}).

\bibitem[{\citenamefont{Forgacs and Mason}(1959{\natexlab{a}})}]{Forgac1959}
\bibinfo{author}{\bibfnamefont{O.}~\bibnamefont{Forgacs}} \bibnamefont{and}
  \bibinfo{author}{\bibfnamefont{S.}~\bibnamefont{Mason}},
  {\bibinfo{title} {\bibnamefont{``Particle motions in sheared suspensions: IX. Spin and
deformation of threadlike particles''}}},
 \bibinfo{journal}{J.  Colloid Sci.} \textbf{\bibinfo{volume}{14}}, \bibinfo{pages}{457--472}
  (\bibinfo{year}{1959}{\natexlab{a}}).

\bibitem[{\citenamefont{Forgacs and Mason}(1959{\natexlab{b}})}]{Forgac1959b}
\bibinfo{author}{\bibfnamefont{O.}~\bibnamefont{Forgacs}} \bibnamefont{and}
  \bibinfo{author}{\bibfnamefont{S.}~\bibnamefont{Mason}},
  {\bibinfo{title} {\bibnamefont{``Particle motions in sheared suspensions: X. Orbits of
 flexible thread-like particles ''}}},
  \bibinfo{journal}{J. Colloid Sci.} \textbf{\bibinfo{volume}{14}}, \bibinfo{pages}{473--491}
  (\bibinfo{year}{1959}{\natexlab{b}}).

\bibitem[{\citenamefont{Leal}(1975)}]{Leal1975}
\bibinfo{author}{\bibfnamefont{L.}~\bibnamefont{Leal}},
 {\bibinfo{title} {\bibnamefont{``The slow motion of slender rod-like particles
  in a second-order fluid''}}},
\bibinfo{journal}{J. Fluid Mech.} \textbf{\bibinfo{volume}{69}}, \bibinfo{pages}{305--337}
  (\bibinfo{year}{1975}).

\bibitem[{\citenamefont{Habibi et al.}(2007)}]{habibi07}
\bibinfo{author}{\bibfnamefont{M.}~\bibnamefont{Habibi}},
\bibinfo{author}{\bibfnamefont{N. M.}~\bibnamefont{Ribe}},
\bibinfo{author}{\bibfnamefont{D.}~\bibnamefont{Bonn}},
 {\bibinfo{title} {\bibnamefont{``Coiling of elastic ropes''}}},
\bibinfo{journal}{Phys. Rev. Lett.} \textbf{\bibinfo{volume}{99}}, \bibinfo{pages}{154302}
  (\bibinfo{year}{2007}).

\bibitem[{\citenamefont{Landau and Lifshitz}(1986)}]{Landau}
\bibinfo{author}{\bibfnamefont{L.}~\bibnamefont{Landau}} \bibnamefont{and}
  \bibinfo{author}{\bibfnamefont{E.}~\bibnamefont{Lifshitz}},
  \emph{\bibinfo{title}{Theory of Elasticity, Third Edition: Volume 7}}
  (\bibinfo{publisher}{Butterworth-Heinemann}, \bibinfo{year}{1986}).

\bibitem[{\citenamefont{Boschan et~al.}(2007)\citenamefont{Boschan, Auradou,
  Ippolito, Chertcoff, and Hulin}}]{Boschan2006}
\bibinfo{author}{\bibfnamefont{A.}~\bibnamefont{Boschan}},
  \bibinfo{author}{\bibfnamefont{H.}~\bibnamefont{Auradou}},
  \bibinfo{author}{\bibfnamefont{I.}~\bibnamefont{Ippolito}},
  \bibinfo{author}{\bibfnamefont{R.}~\bibnamefont{Chertcoff}},
  \bibnamefont{and} \bibinfo{author}{\bibfnamefont{J.}~\bibnamefont{Hulin}},
  {\bibinfo{title} {\bibnamefont{``Miscible displacement fronts of shear thinning fluids
inside rough fractures''}}},
  \bibinfo{journal}{Water Res. Res.} \textbf{\bibinfo{volume}{43}},
  \bibinfo{pages}{W03438} (\bibinfo{year}{2007}).

\bibitem[{\citenamefont{{Boffa} et~al.}(1998)\citenamefont{{Boffa}, {Allain},
  and {Hulin}}}]{Boffa1998}
\bibinfo{author}{\bibfnamefont{J.~M.} \bibnamefont{{Boffa}}},
  \bibinfo{author}{\bibfnamefont{C.}~\bibnamefont{{Allain}}}, \bibnamefont{and}
 \bibinfo{author}{\bibfnamefont{J.~P.} \bibnamefont{{Hulin}}},
 {\bibinfo{title} {\bibnamefont{``Experimental analysis of fracture rugosity
in granular and compact rocks''}}},
 \bibinfo{journal}{European Physical Journal Applied Physics}
 \textbf{\bibinfo{volume}{2}}, \bibinfo{pages}{281--289} (\bibinfo{year}{1998}).

\bibitem[{\citenamefont{Bouchaud}(2003)}]{Bouchaud2003}
\bibinfo{author}{\bibfnamefont{E.}~\bibnamefont{Bouchaud}},
 {\bibinfo{title} {\bibnamefont{``The morphology of fracture surfaces,
 a tool to understand crack propagation in complex materials''}}},
  \bibinfo{journal}{Surf. Rev. Lett.} \textbf{\bibinfo{volume}{10}},
  \bibinfo{pages}{797-814} (\bibinfo{year}{2003}).



\bibitem[{\citenamefont{Poon et~al.}(1992)\citenamefont{Poon, Sayles, and
  Jones}}]{Poon1992}
\bibinfo{author}{\bibfnamefont{C.}~\bibnamefont{Poon}},
  \bibinfo{author}{\bibfnamefont{R.}~\bibnamefont{Sayles}}, \bibnamefont{and}
  \bibinfo{author}{\bibfnamefont{T.}~\bibnamefont{Jones}},
   {\bibinfo{title} {\bibnamefont{``Surface measurement and fractal characterization of naturally fractured rocks''}}},
   \bibinfo{journal}{J. Phys. D: Appl. Phys.} \textbf{\bibinfo{volume}{25}}, \bibinfo{pages}{1269--75}
  (\bibinfo{year}{1992}).

\bibitem[{\citenamefont{D'Angelo et~al.}(2007)\citenamefont{D'Angelo, Auradou, Allain, and
  Hulin}}]{Dangelo2007}
\bibinfo{author}{\bibfnamefont{M.V.}~\bibnamefont{D'Angelo}},
  \bibinfo{author}{\bibfnamefont{H.}~\bibnamefont{Auradou}},
  \bibinfo{author}{\bibfnamefont{C.}~\bibnamefont{Allain}}, \bibnamefont{and}
  \bibinfo{author}{\bibfnamefont{J.P.}~\bibnamefont{Hulin}},
  {\bibinfo{title} {\bibnamefont{``Pore scale mixing and macroscopic solute dispersion regimes in polymer
flows inside two-dimensional model networks''}}},
  \bibinfo{journal}{Phys. Fluids} \textbf{\bibinfo{volume}{19}},
  \bibinfo{pages}{033103} (\bibinfo{year}{2007}).

\bibitem[{\citenamefont{Zimmerman et~al.}(2004)\citenamefont{Zimmerman,
  Al-Yaarubi, Pain, and Grattoni}}]{Zimmerman2004}
\bibinfo{author}{\bibfnamefont{W.}~\bibnamefont{Zimmerman}},
  \bibinfo{author}{\bibfnamefont{A.}~\bibnamefont{Al-Yaarubi}},
  \bibinfo{author}{\bibfnamefont{C.}~\bibnamefont{Pain}}, \bibnamefont{and}
  \bibinfo{author}{\bibfnamefont{C.}~\bibnamefont{Grattoni}},
  {\bibinfo{title} {\bibnamefont{``Non-linear regimes of fluid flow in rock fractures''}}},
  \bibinfo{journal}{Int. J. Rock Mech. Min. Sci.}
  \textbf{\bibinfo{volume}{41}}, \bibinfo{pages}{163--169} (\bibinfo{year}{2004}).

\bibitem[{\citenamefont{Gabbanelli et~al.}(2005)\citenamefont{Gabbanelli,
  Drazer, and Koplik}}]{Gabbanelli2005}
\bibinfo{author}{\bibfnamefont{S.}~\bibnamefont{Gabbanelli}},
  \bibinfo{author}{\bibfnamefont{G.}~\bibnamefont{Drazer}}, \bibnamefont{and}
  \bibinfo{author}{\bibfnamefont{J.}~\bibnamefont{Koplik}},
  {\bibinfo{title} {\bibnamefont{``Lattice Boltzmann method for non-
Newtonian (power-law) fluids''}}},
  \bibinfo{journal}{Phys. Rev. E.} \textbf{\bibinfo{volume}{72}},
  \bibinfo{pages}{046312} (\bibinfo{year}{2005}).

 \bibitem[{\citenamefont{Gentier et~al.}(1997)\citenamefont{Gentier, Lamontagne,
  Archambault, and Riss}}]{Gentier1997}
\bibinfo{author}{\bibfnamefont{S.}~\bibnamefont{Gentier}},
  \bibinfo{author}{\bibfnamefont{E.}~\bibnamefont{Lamontagne}},
  \bibinfo{author}{\bibfnamefont{G.}~\bibnamefont{Archambault}},
  \bibnamefont{and} \bibinfo{author}{\bibfnamefont{J.}~\bibnamefont{Riss}},
  {\bibinfo{title} {\bibnamefont{``Anisotropy of flow in a
fracture undergoing shear and its relationship to the direction of
shearing and injection pressure''}}},
  \bibinfo{journal}{Int. J. Rock Mech. Min. Sci.}
  \textbf{\bibinfo{volume}{34}}, \bibinfo{pages}{412--412} (\bibinfo{year}{1997}).

\bibitem[{\citenamefont{Yeo et~al.}(1998)\citenamefont{Yeo, Freitas, and
  Zimmerman}}]{Yeo1998}
\bibinfo{author}{\bibfnamefont{I.~W.} \bibnamefont{Yeo}},
  \bibinfo{author}{\bibfnamefont{M.~H.~D.} \bibnamefont{Freitas}},
  \bibnamefont{and} \bibinfo{author}{\bibfnamefont{R.~W.}
  \bibnamefont{Zimmerman}},
   {\bibinfo{title} {\bibnamefont{``Effect of shear displacement on the
aperture and permeability of a rock fracture''}}},
   \bibinfo{journal}{Int. J. Rock Mech. Min. Sci.}
  \textbf{\bibinfo{volume}{35}}, \bibinfo{pages}{1051--1070} (\bibinfo{year}{1998}).

\bibitem[{\citenamefont{Auradou et~al.}(2006)\citenamefont{Auradou, Drazer,
  Boschan, Hulin, and Koplik}}]{Auradou2006}
\bibinfo{author}{\bibfnamefont{H.}~\bibnamefont{Auradou}},
  \bibinfo{author}{\bibfnamefont{G.}~\bibnamefont{Drazer}},
  \bibinfo{author}{\bibfnamefont{A.}~\bibnamefont{Boschan}},
  \bibinfo{author}{\bibfnamefont{J.-P.} \bibnamefont{Hulin}}, \bibnamefont{and}
  \bibinfo{author}{\bibfnamefont{J.}~\bibnamefont{Koplik}},
  {\bibinfo{title} {\bibnamefont{``Flow channeling in a single fracture induced
by shear displacement''}}},  \bibinfo{journal}{Geothermics}
\textbf{\bibinfo{volume}{35}},
  \bibinfo{pages}{576--588} (\bibinfo{year}{2006}).

\bibitem[{\citenamefont{Auradou et~al.}(2008)\citenamefont{Auradou, Boschan, Chertcoff, Gabbanelli, Hulin, and
  Ippolito}}]{Auradou2008}
\bibinfo{author}{\bibfnamefont{H.}~\bibnamefont{Auradou}},
\bibinfo{author}{\bibfnamefont{A.}~\bibnamefont{Boschan}},
\bibinfo{author}{\bibfnamefont{R.}~\bibnamefont{Chertcoff}},
\bibinfo{author}{\bibfnamefont{S.}~\bibnamefont{Gabbanelli}},
\bibinfo{author}{\bibfnamefont{J.P.}~\bibnamefont{Hulin}}, \bibnamefont{and}
\bibinfo{author}{\bibfnamefont{I.}~\bibnamefont{Ippolito}},
 {\bibinfo{title} {\bibnamefont{``Enhancement of velocity contrasts by shear thinning
solutions flowing in a rough fracture''}}},
\bibinfo{journal}{J. N. Newt. Fluid. Mech.}
 \bibinfo{pages}{doi : 10.1016/j.jnnfm.2007.11.008}  (\bibinfo{year}{2008}).

\end{thebibliography}

\end{document}